\newcount\draft\draft=0 

\documentclass[acmtog]{acmart}

\renewcommand\footnotetextcopyrightpermission[1]{} % removes footnote with conference information in first column
\acmConference[AI4CS]{AI/ML for Cybersecurity: Challenges, Solutions, and Novel
Ideas at SDM '21}{April 2021}{NY}

\usepackage{tikz}
\usepackage{amsmath}

\usepackage{float}
\usepackage{enumerate}

\usepackage{bm}
\usepackage{soul}
\usepackage[normalem]{ulem}

\ifcsname G\endcsname%
\else%
\fi%
\ifcsname C\endcsname%
\else%
\fi%

\renewcommand{\S}{\mathcal{S}}
\newcommand{\A}{\mathcal{A}}

\renewcommand{\P}{\mathcal{P}}

\ifnum\draft>0
  
\else 
  
\fi

\ifnum\draft>1
  \newcommand{\removedtext}[1]{\textcolor{red}{\sout{#1}}}
\else

  \newcommand{\removedtext}[1]{\ignorespaces}
\fi

\usepackage{tabu}

\usepackage{wrapfig}
\usepackage{graphicx}
\usepackage[all]{xy}

\usepackage{color}
\usepackage{fancyvrb}

\usepackage{cleveref}
\crefname{section}{§}{§§}
\Crefname{section}{§}{§§}

\newcommand\blfootnote[1]{%
  \begingroup
  \renewcommand\thefootnote{}\footnote{#1}%
  \addtocounter{footnote}{-1}%
  \endgroup
}

\begin{document}
%%%%%%%%%%%%%%%%%%%%%%%%%%%%%%%%%%%%%%%%%%%%%%%%%%%%%%%%%%%%%%%%%%%%%%%%%%%%%%%%

\date{}

\begin{abstract}
Research seeks to apply Artificial Intelligence (AI) to scale and extend the
capabilities of human operators to defend networks. A fundamental problem that
hinders the generalization of successful AI approaches~--i.e., beating humans at
playing games~--is that network defense cannot be defined as a single game with
a fixed set of rules. Our position is that network defense is better
characterized as a collection of games with uncertain and possibly drifting
rules. Hence, we propose to define network defense tasks as distributions of
\emph{network environments}, to: (i) enable research to apply
modern AI techniques, such as unsupervised curriculum learning and reinforcement
learning for network defense; and, (ii) facilitate the design of
\emph{well-defined challenges} that can be used to compare approaches for
autonomous cyberdefense. 

To demonstrate that an approach for autonomous network defense is practical it
is important to be able to reason about the boundaries of its applicability.
Hence, we need to be able to define network defense tasks that capture sets of
adversarial tactics, techniques, and procedures (TTPs); quality of service (QoS)
requirements; and TTPs available to defenders. Furthermore, the abstractions to
define these tasks must be extensible; must be backed by well-defined semantics
that allow us to reason about distributions of environments; and should enable
the generation of data and experiences from which an agent can learn.

We outline key aspects of network environments that must be parametrized, and we
motivate the use of \emph{generative programs} to define probability
distributions over network environments. We explain how this approach can enable
the development of the next-generation of autonomous cyberdefenders, which will
not only have to face adversaries that target the network, but also the
AI-enabled components of cyberdefense~--by automating attacks and by morphing
behavior in systematic ways to poison or evade defender models. Our approach
named \emph{Network Environment Design for Autonomous Cyberdefense} inspired the
architecture of FARLAND, a Framework for Advanced Reinforcement Learning for
Autonomous Network Defense, which we use at MITRE to develop RL network
defenders that perform blue actions from the MITRE Shield matrix against
attackers with TTPs that drift from MITRE ATT\&CK TTPs.

\end{abstract}

\title{Network Defense is Not a Game}

\author{Andres Molina-Markham}
\email{a.mm@mitre.org}
\author{Ransom K. Winder}
\affiliation{%
  \institution{The MITRE Corporation}
}

\author{Ahmad Ridley}
\affiliation{%
  \institution{National Security Agency}}

\maketitle

\blfootnote{Approved for Public Release; Distribution Unlimited. Public Release Case Number 21-0464. 
This technical data deliverable was developed using contract funds under Basic
Contract No. W56KGU-18-D-0004. The view, opinions, and/or findings contained in
this report are those of The MITRE Corporation and should not be construed as an
official Government position, policy, or decision, unless designated by other
documentation. \copyright 2021 The MITRE Corporation. ALL RIGHTS RESERVED.}

%-------------------------------------------------------------------------------

\section{Introduction}

One fundamental problem in applying Reinforcement Learning (RL) to network defense
is that network defense cannot be defined as a single game with a simple set of
rules. Rather, proficient network defense corresponds to mastering a spectrum of
games that depend on sets of (i) adversarial tactics, techniques, and procedures
(TTPs); (ii) quality of service goals and characteristics of a network; and
(iii) actions available for defenders and their security goals. 

In this paper, we argue that it is necessary to define network defense tasks via
distributions of network environments to facilitate the development  and
evaluation of RL approaches for autonomous cyberdefense. Our position is based on
our work~\cite{molina-markham_network_2021}, which proposes: (a) aspects of
network models that must be available to researchers seeking to apply RL to
reconfigure networks to mitigate cyberattacks; and (b) a way to use generative
programs to define distributions of network environments.

Network environment design~\cite{molina-markham_network_2021} directly addresses
the problem of enabling progress toward learning a complex task. Furthermore, it
addresses the more fundamental problem that network defense cannot be defined as
a single game with a simple set of rules. This is key when applying RL to
solving a problem that changes over time. Autonomous network defenders must not
only be able to reconfigure hosts and networks to mitigate common adversaries
such as those described in MITRE’s ATT\&CK framework~\cite{strom_mitre_2018}.
Rather, defenders must increasingly be concerned about more sophisticated
adversaries that target autonomous cyber defense. Instead of demonstrating RL's
effectiveness for autonomous cyberdefense, our work illustrates the perils to
the defender of
adversaries capable of deception and indirect manipulation of observations the
network defense uses. 

This approach inspired
FARLAND~\cite{molina-markham_network_2021} (a
Framework for Advanced Reinforcement Learning for Autonomous Network Defense), 
a research collaboration project between MITRE and NSA,
which allows for the development of robust RL network defenders on simulated or
emulated networks. Unlike RL systems and frameworks
(~\cite{brockman_openai_2016, beattie_deepmind_2016, johnson_malmo_2016}),
FARLAND (i) can be used to guide an agent through a sequence of network defense
tasks of increasing difficulty to guarantee progress, and (ii) exposes a set of
model parameters specific to address the problem of network defense.
 
\section{The network environment design task for autonomous cyberdefense}

RL agents learn to execute complex tasks through observing the effects of their
actions on an environment. Observations consist of state features and rewards.
Rewards are quantities that encode the desirability of the effects. The goal of
an RL-agent is to learn to act to maximize the sum of expected rewards over time.

When developing RL agents, it is common to use a game abstraction to frame the
learning task. However, as we noted in our \emph{Network Environment Design for
Autonomous Cyberdefense} approach~\cite{molina-markham_network_2021}, the
practicality of a network defense policy must not be evaluated based on the
performance of an agent under predominantly fixed configurations. Cyberdefender
agents (blue agents) should not simply attempt to defeat an adversary (red
agent) with fixed TTPs. Instead they must account for changes in behaviors that
effectively change the game itself, in particular with respect to the red agent
TTPs. The range of red agent behaviors should not be arbitrary either. That
would make it impossible to learn and evaluate policies. This section summarizes
our key ideas about modeling distributions of \emph{network environments} using
generative programs~\cite{molina-markham_network_2021}.
 
For a fixed network environment, the goal of a blue agent is to
learn to achieve a high score by performing actions on a network to maintain an
acceptable level of service for authorized hosts subject to a resource budget
and while preventing unauthorized access. Thus, the learning agent must master
an operator role in a cybersecurity operation center (CSOC). Network defense
tasks should be specified over ranges of parameters, or more precisely, over
distributions of network environments. By following this approach, it is
possible to reason about the performance boundaries of an agent’s policy. In
turn, such ability to reason about an agent's performance can inform strategies
to gradually increase the complexity of tasks to facilitate progress; or to
divide the network defense problem into smaller problems to apply hierarchical
approaches~\cite{wen_efficiency_2020}. Our approach is compatible with related
ideas, including Automated Domain Randomization~\cite{openai_solving_2019}, and
Unsupervised Environment Design~\cite{dennis_emergent_2020}. 

In this section we enumerate key aspects of a network environment that must
be parametrizable, and we outline our approach~\cite{molina-markham_network_2021}
for modeling distributions of network environments using generative
programs~\cite{cusumano-towner_gen:_2019}.
 
\subsection{Features of network environments}
\label[]{subsec:features}

Network environment definitions include the definition of action spaces,
observation spaces, and reward functions, as well as the definition of
probabilistic models that characterize network dynamics (including how devices
are connected) and the behaviors of adversarial (red) and benign (gray) agents.
Because all these aspects can vary dramatically, the \emph{network environment
design} task is to delimit them in a way that is conducive to developing
and evaluating learning approaches.

\textbf{Agent behavior.} Agents (red, gray, and blue) take actions that alter
the network state, but each agent only partially and imperfectly observes the
state of the network. Hence, defining the behavior of agents requires
definitions of actions (in $\A$) that affect the network, that agents take based
on observations (in $\Omega$), which depend on the state of the network (in
$\S$). In our approach~\cite{molina-markham_network_2021}, the process of
determining future actions (the policy) is provided via generative programs in
the case of red and gray agents. Blue agents learn policies via RL algorithms.
Different agents may estimate states differently, aware that each has different
goals and observability capabilities. 

\textbf{Game state.} The state of a network depends on how devices are connected
and how network configurations change due to agents' actions. Typical network
configurations include a set of hosts, which may change during an episode. Blue
agents may have full visibility about which nodes are connected on a network and
how; services that are running in each host and which ports are used;
information about which hosts contain the crown jewels to protect; and packet
forwarding rules. In contrast, red agents only have partial and imperfect
information about hosts and network configurations. Our paper discusses how
FARLAND~\cite{molina-markham_network_2021} maintains state using graphs, and how
portions of the state are visible to red, gray, and blue agents. 

\textbf{Observations.} Because it is impractical to assume that agents will be
able to fully observe the network state, agents act on partial and imperfect
state information. Part of a network environment model is to determine what
features of the state are observable by which agent and with what kind of
uncertainty. One may assume that gray agents act on information available to
normal authorized users. Blue agents may have access to much more information,
but it may be necessary to only focus on a subset of the state space for the
purposes of making the learning process feasible. For example, a blue agent may
take actions based on observations that summarize the state of the
environment at a snapshot in time, summarizing network information and host
activity information primarily related to quality of service and threat
indicators. Network information may include specific events (e.g., host A sent a
file to host B via scp), or they may include network statistics that, for
example, describe volumes of traffic in network regions or between specific
hosts. Host activity information describes events observed by monitors on the
hosts and do not necessarily involve network events or generation of traffic. 

\textbf{Rewards.} Reward functions are specified by the researcher and take as
inputs features of the network state; costs associated with deploying or
repairing services; and potentially subjective representations of how good or
bad certain states are for a set of security goals. While these may be
subjective, because different organizations may value security goals
differently, broadly speaking, indications of successful compromises or service
degradation should incur penalties. Successful attack containment and adequate
service would result in positive rewards. For a given environment, the set of
suitable reward functions is not unique.

\subsection{Generative programs to model network environment distributions}

Our discussion in Section~\ref{subsec:features} enumerates several aspects of a
network environment that must be specified to define a network
defense task. The next question is: how do we represent distributions of network
environments in a manner that allows researchers to develop novel RL approaches
for network defense? Our
work~\cite{molina-markham_network_2021} proposed the
use of generative programs to represent distributions of network environments. 

Generative programs define probability distributions on execution
traces~\cite{cusumano-towner_gen:_2019}. Instead of defining deterministic
mappings between inputs and outputs, they define a weighted set $\{(x, \xi)\}$
of possible execution traces, where $x$ denotes a trace, and $\xi$ denotes its
weight, associated with the probability that running the corresponding program
$\P$ with specific arguments $\alpha$, results in $x$. In our case, generative
programs allow researchers to model stochastic aspects of network behavior, such
as gray agents or failures on a network. Additionally, they allow the
specification of assumptions about red TTP variability for deception purposes. 

Parameters of generative programs can be learned from data. This allows, for
example, the generation of synthetic behavior that captures observations from real
logs and network traffic. In contrast to using generative adversarial networks
(GANs)~\cite{goodfellow_generative_2014}, which can generate synthetic data
based on datasets, generative programs facilitate the implementation of planners
with stochastic behavior. In particular, the structure of a red TTP can be
described by a program that resembles a traditional implementation (e.g., an
automated planner) but during execution, red and gray agents can make choices
determined by probability distributions, which could result in dramatically
different traces.

A key observation for network defense is that a game requires maintaining QoS,
preventing a red agent from achieving its goals, and recognizing potentially
deceptive behaviors on the part of the red agent. Therefore, if the behaviors
are sufficiently complex and non-deterministic, the blue agent must learn to
succeed at more than one simple game.

\subsection{Motivating example: poisoning an RL network defender}

We have performed initial experiments leveraging the approach we describe above.
Our results illustrate that, when we train a blue agent using known TTPs without
accounting for potential (poisoning or evasion) attacks targeting the learning
algorithm, its performance cannot be  guaranteed or generalized. In prior work
we showed that while a blue agent may achieve acceptable performance when
trained against a red agent that is only assumed to behave in a manner
consistent with a subset of behaviors described in the ATT\&CK framework, when
we perturb the behavior of the agent using only gray-like actions, the
performance of the blue agent is significantly
degraded~\cite{molina-markham_network_2021}. In other words, blue agents trained
through RL are vulnerable to simple deception. This is because they are playing
only one game without accounting for \emph{``rules''} where the blue agent
itself is targeted.

\begin{figure*}[t!]
    \centering
\begin{tabular}{ccc}
    \includegraphics[width=.3\textwidth,height=3.4cm]{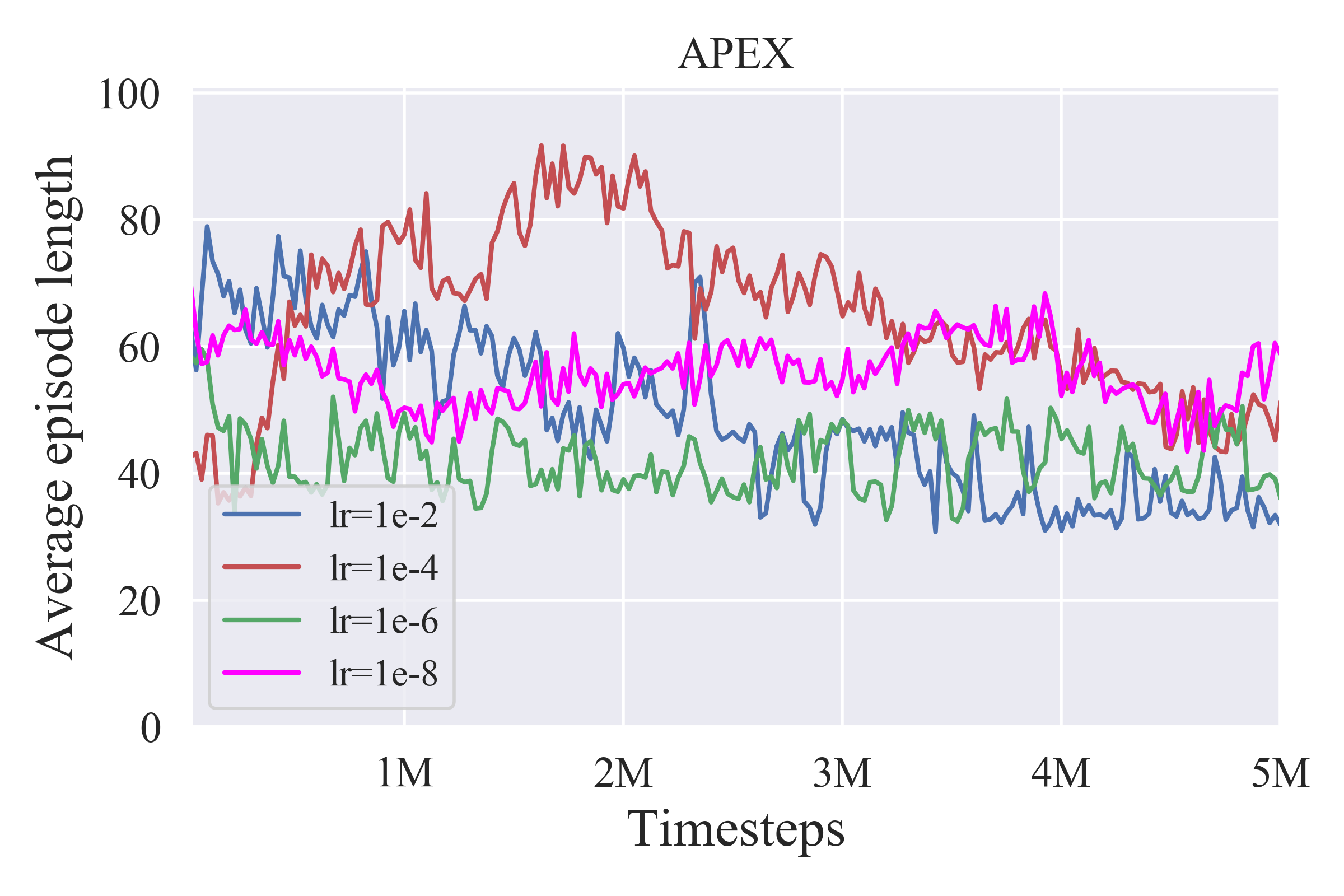} &
    \includegraphics[width=.3\textwidth,height=3.4cm]{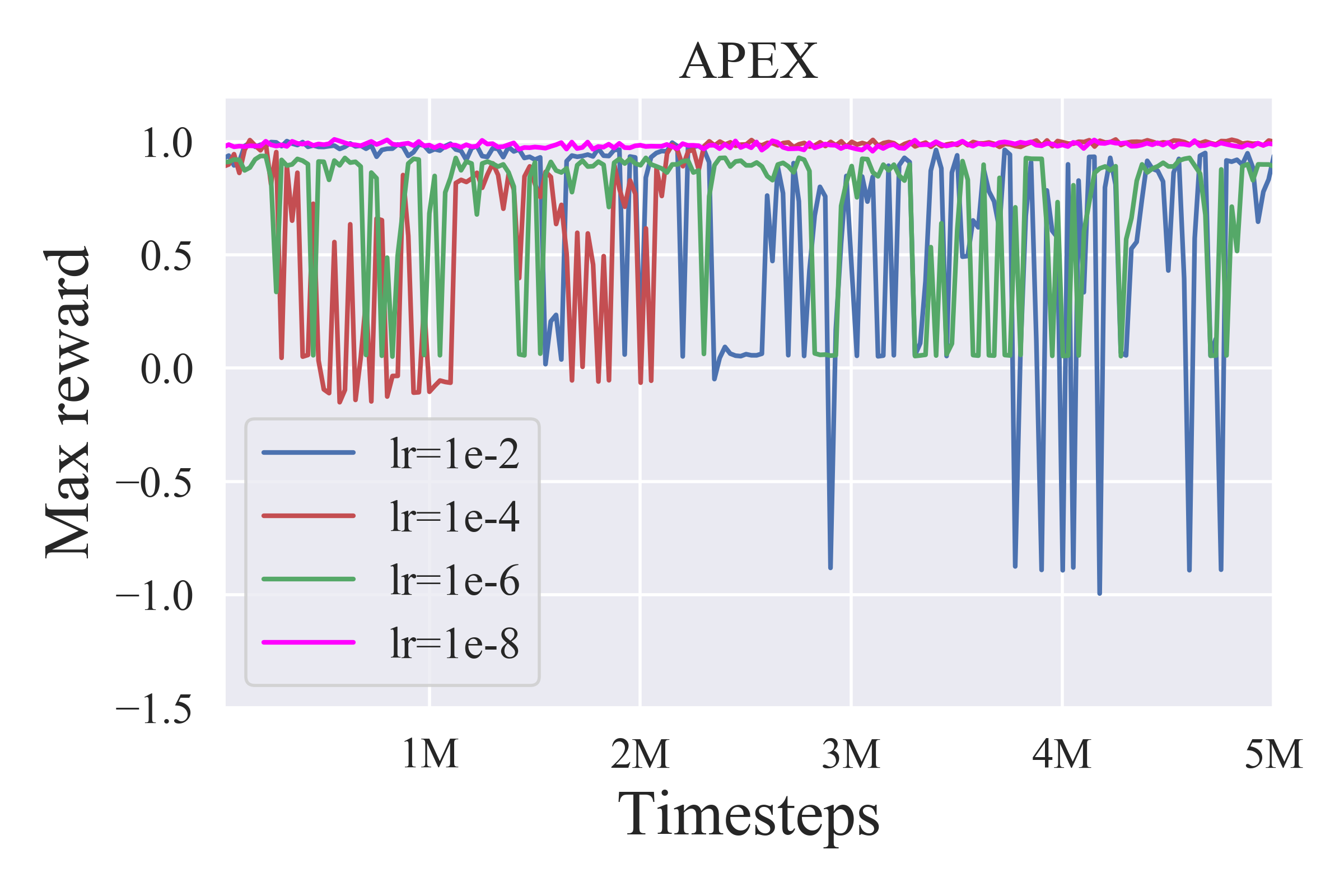} &
    \includegraphics[width=.3\textwidth,height=3.4cm]{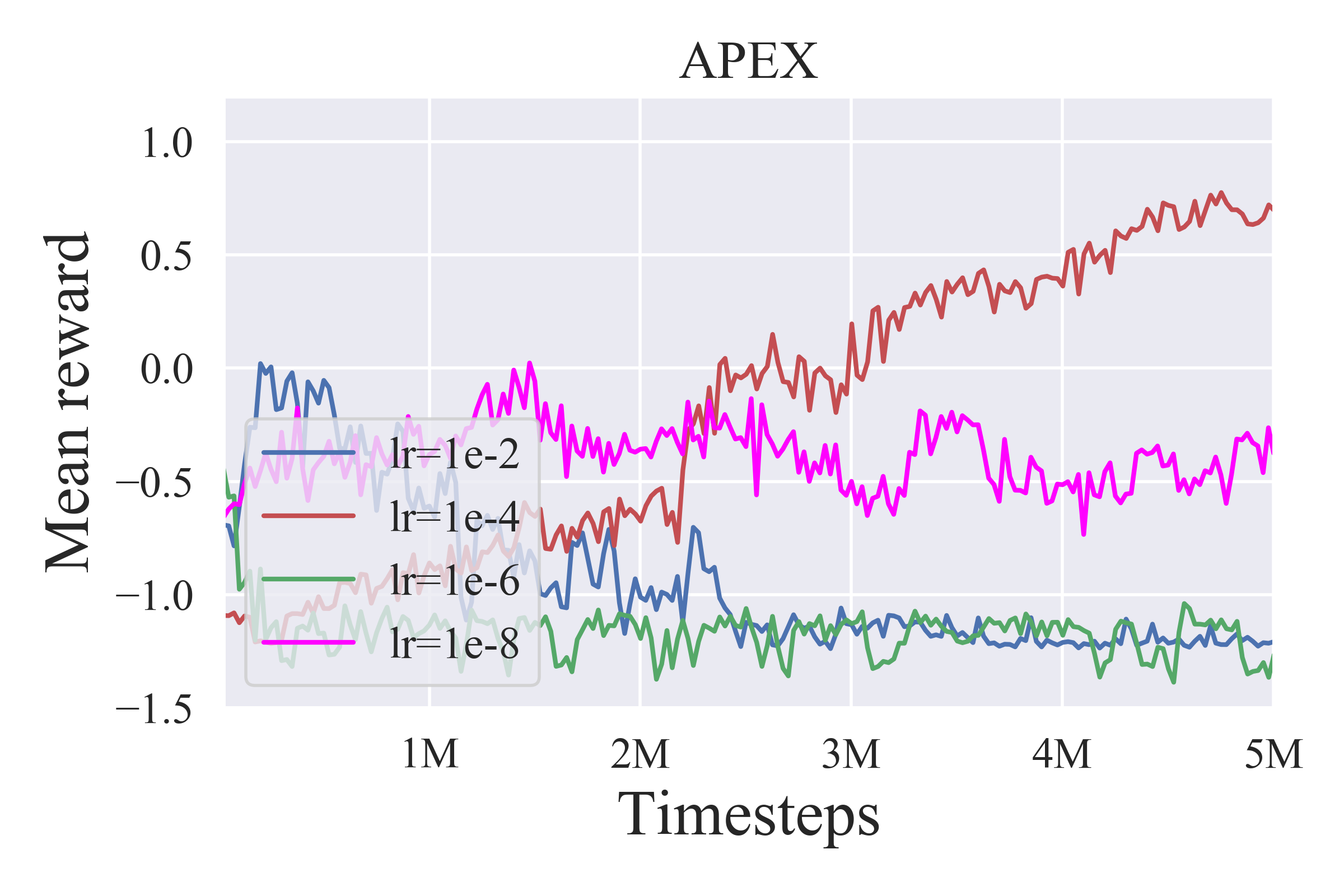} \\
    \includegraphics[width=.3\textwidth,height=3.4cm]{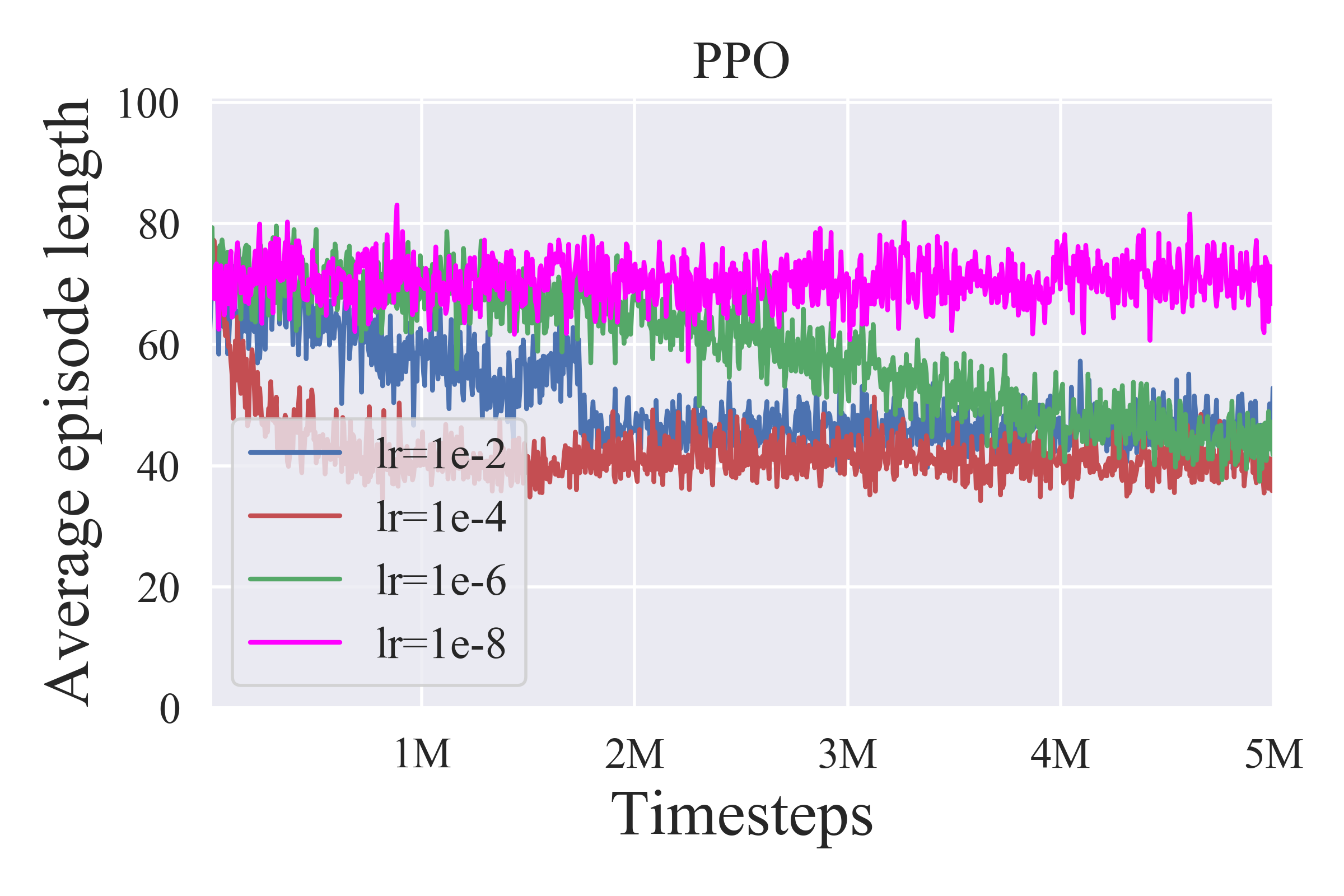} &
    \includegraphics[width=.3\textwidth,height=3.4cm]{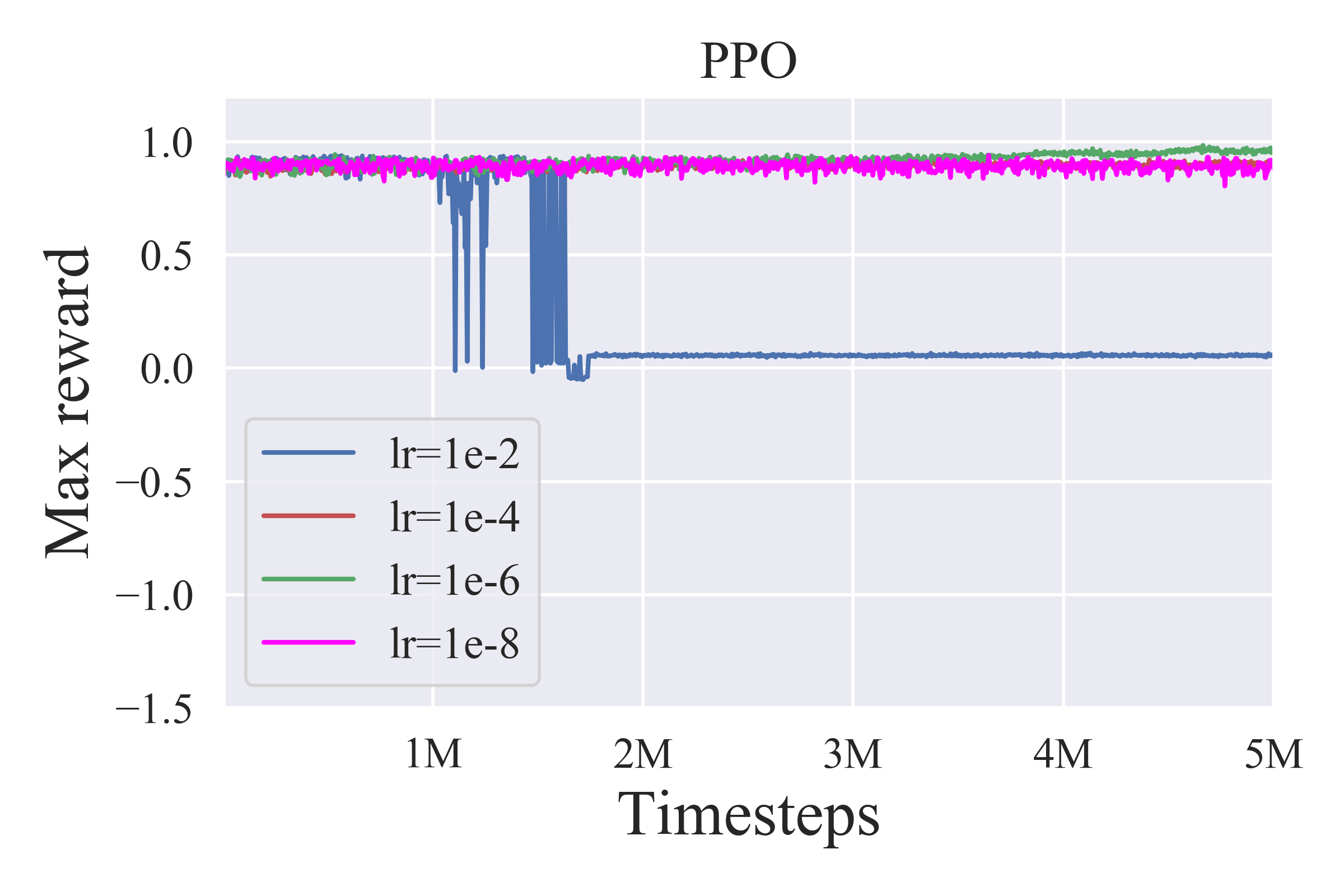} &
    \includegraphics[width=.3\textwidth,height=3.4cm]{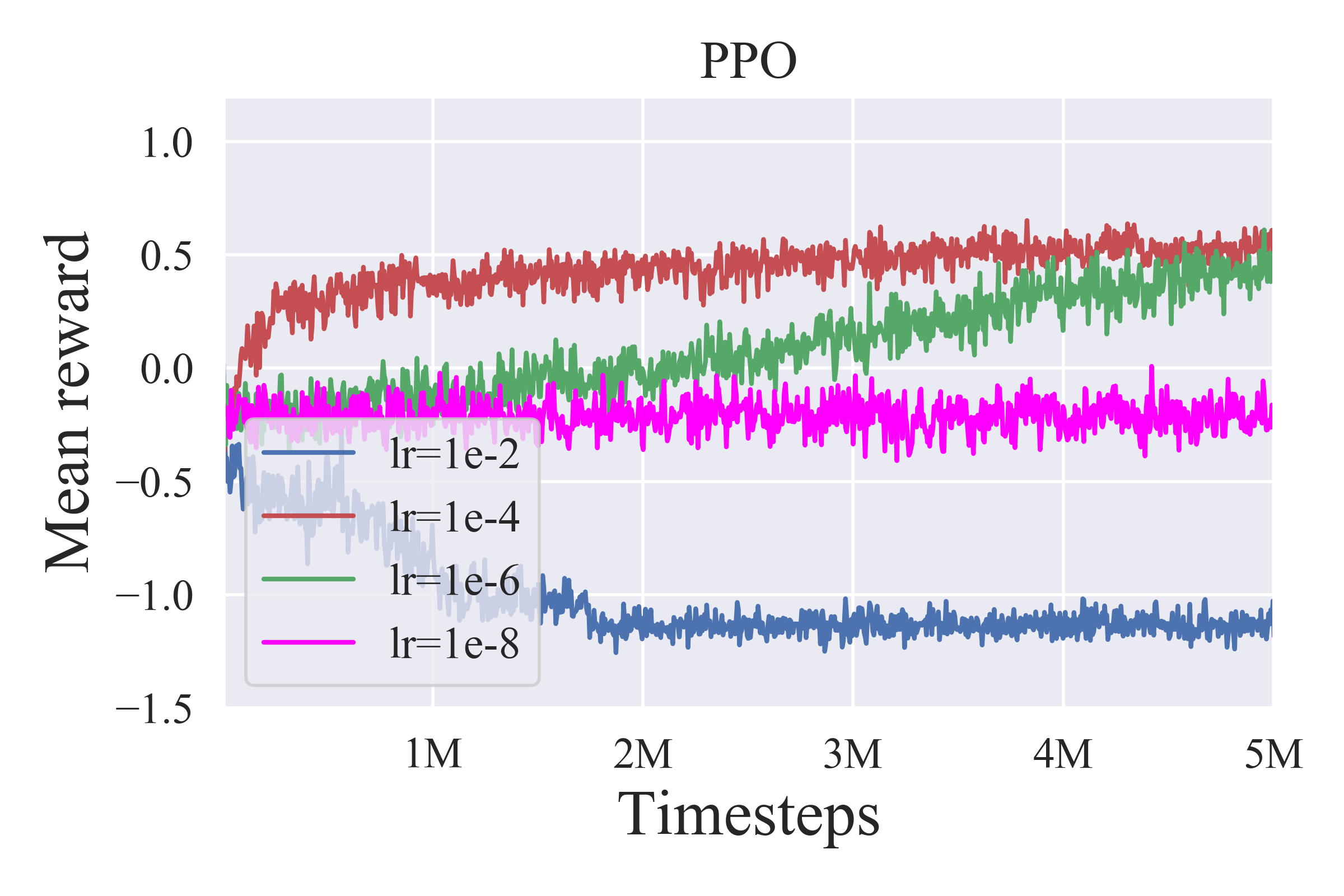} \\
    \includegraphics[width=.3\textwidth,height=3.4cm]{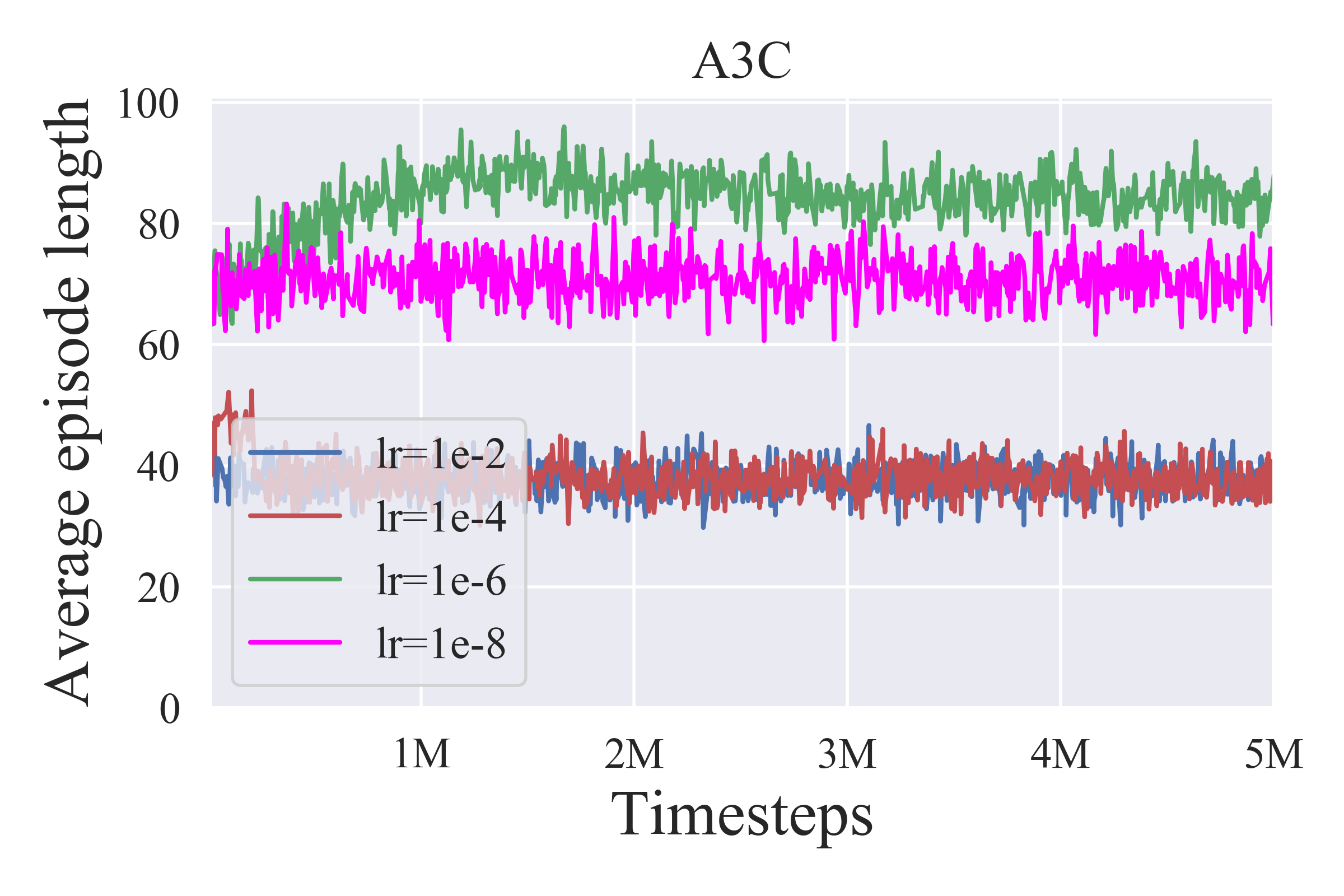} &
    \includegraphics[width=.3\textwidth,height=3.4cm]{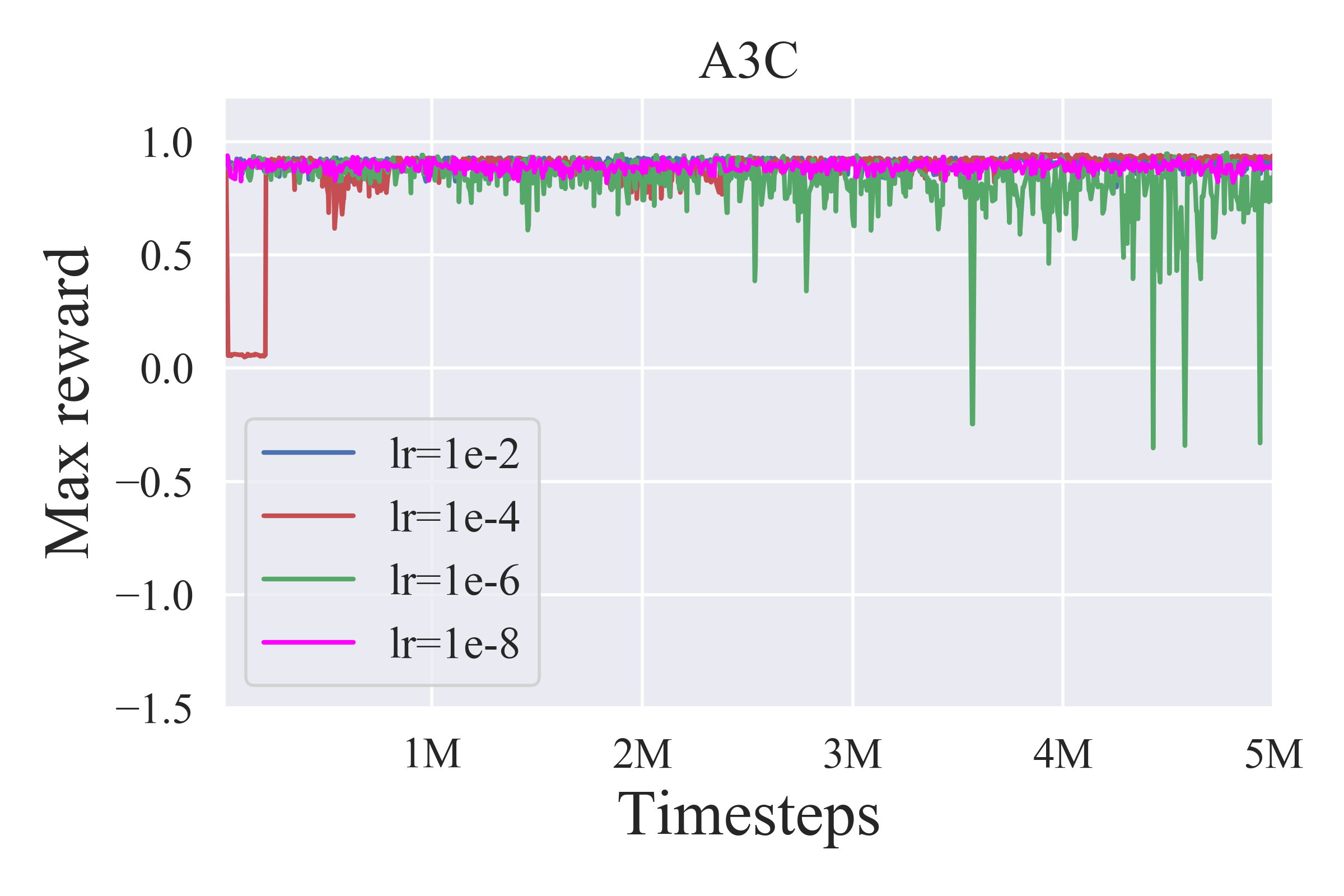} &
    \includegraphics[width=.3\textwidth,height=3.4cm]{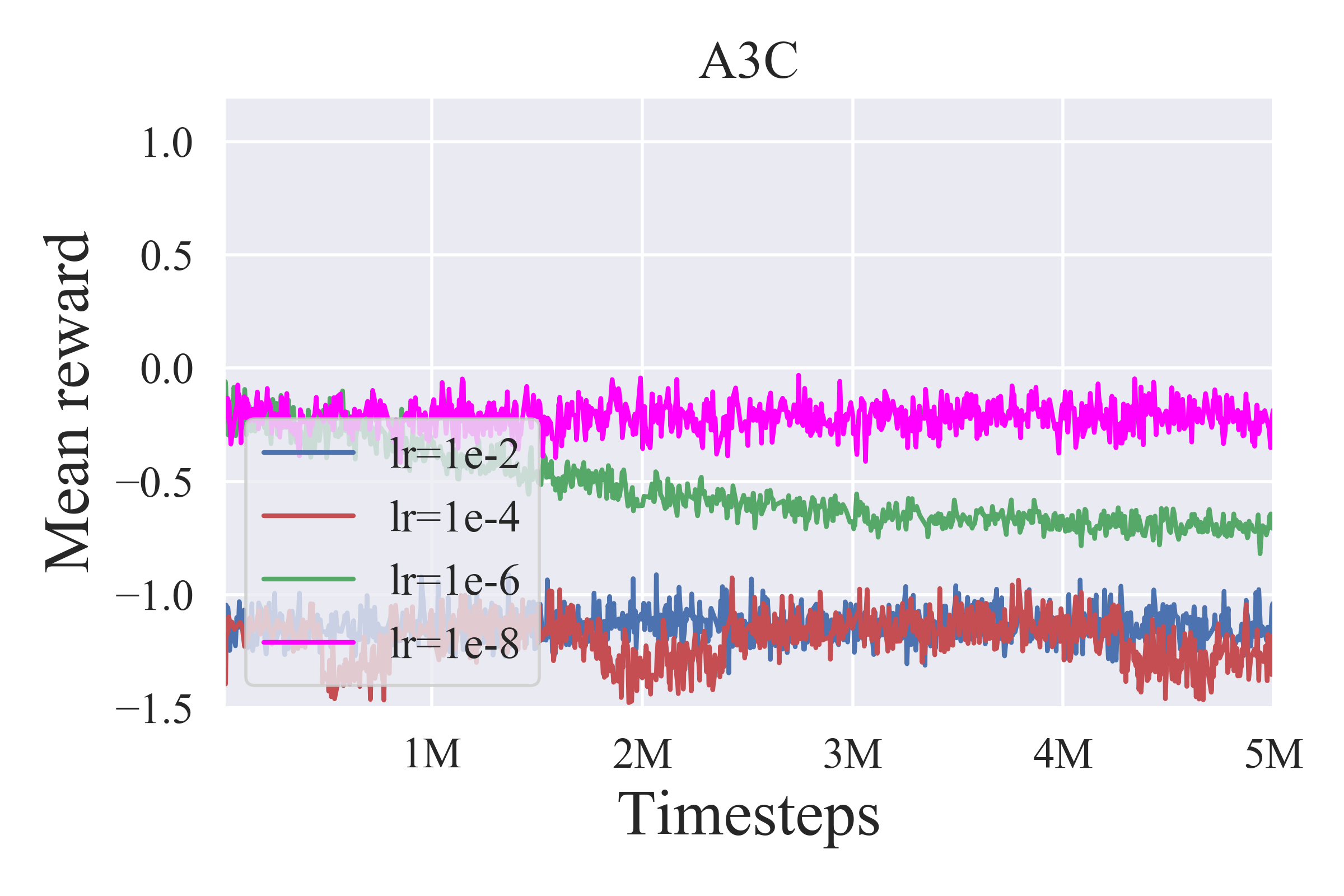} 
\end{tabular}
\caption{Blue agent's performance in a 10-node network against an exfiltration
attack (consistent with ATT\&CK) with three different algorithms,
APEX-DQN~\cite{horgan_distributed_2018},
PPO~\cite{schulman_proximal_2017}, and
A3C~\cite{mnih_asynchronous_2016}. An episode ends when the
red agent exfiltrates a real or a fake crown jewel; or after 100 steps. The left
column shows that all three algorithms result in behavior that lets the red
agent "win". However some learn to let the red agent win in a fake network
rather than in a real network. In the former case, the average reward approaches
1 (e.g., when using APEX-DQN with learning rate lr=0.0001). Other
configurations, (e.g., PPO with lr=0.01) converge to a policy with poor
performance that lets the red agent exfiltrate most of the time. In this case,
5M timesteps correspond to approximately 120K episodes.}
\label{fig:traditional_attack}
\end{figure*}

Figure~\ref{fig:traditional_attack} shows our previous
results~\cite{molina-markham_network_2021} where a blue agent learns with three
different algorithms using various learning rates. The blue agent experiences
an adversary’s TTP (exfiltration) consistent with ATT\&CK, in a 10-node network.
The blue agent’s action space consists of selecting a host and choosing whether
to isolate it, migrate it to an existing network, or migrate it to a new
network. The reward function favors trapping an offender in a honey network over
isolating it. The reward function also penalizes unnecessarily isolating or
migrating hosts to a honey network when they are not misbehaving. The agent
learns to identify characteristics of attacks and QoS from the following
information: the numbers of SCP events, HTTP events, AMQ events, SSH events,
quiet or aggressive reconnaissance events originated at a host; and the numbers of
SCP failures, REST request failures, AMQP failures, SSH failures, and content
searches reported by hosts. 

\begin{figure*}[t!]
    \centering
\begin{tabular}{ccc}
    \includegraphics[width=.3\textwidth]{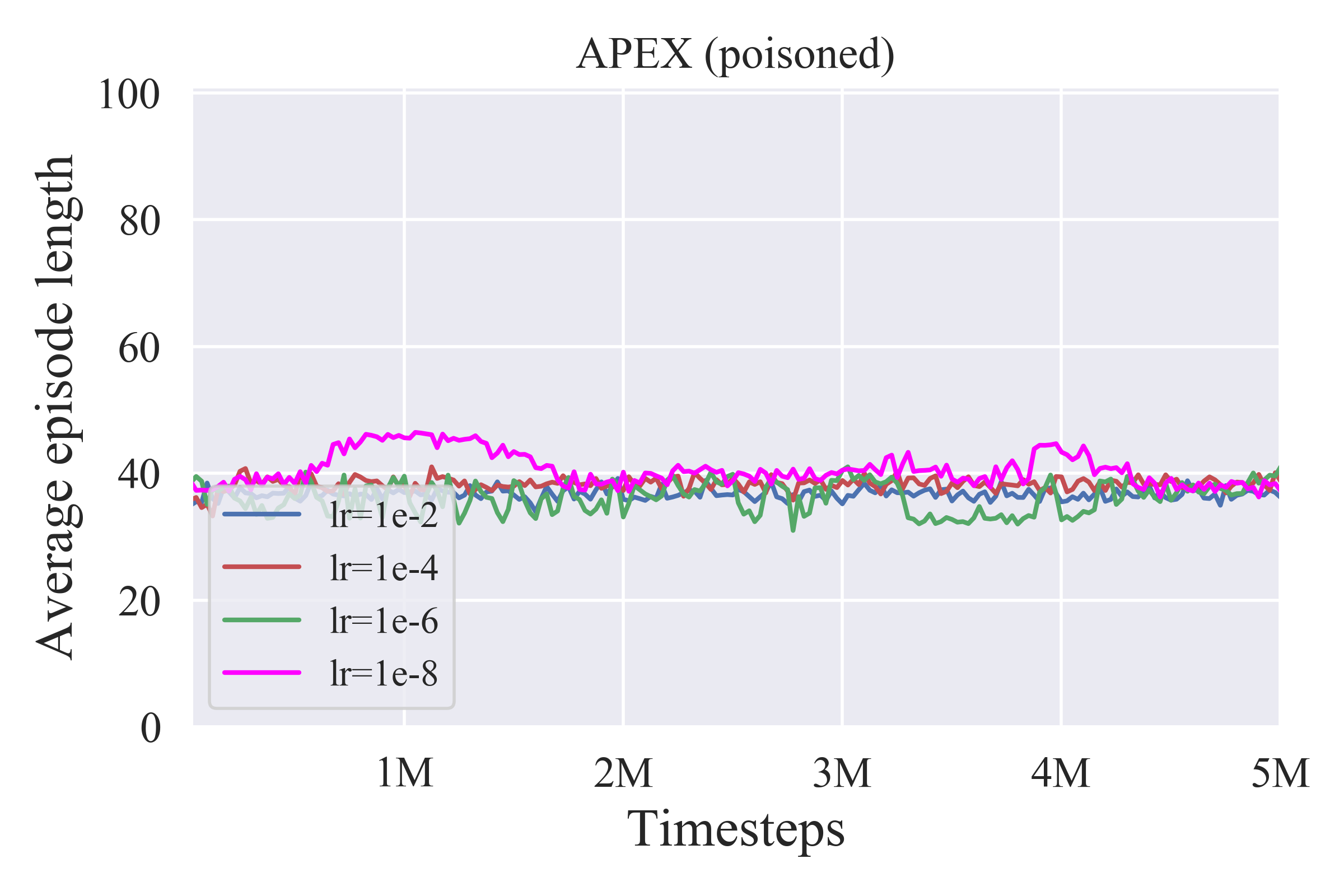} &
    \includegraphics[width=.3\textwidth]{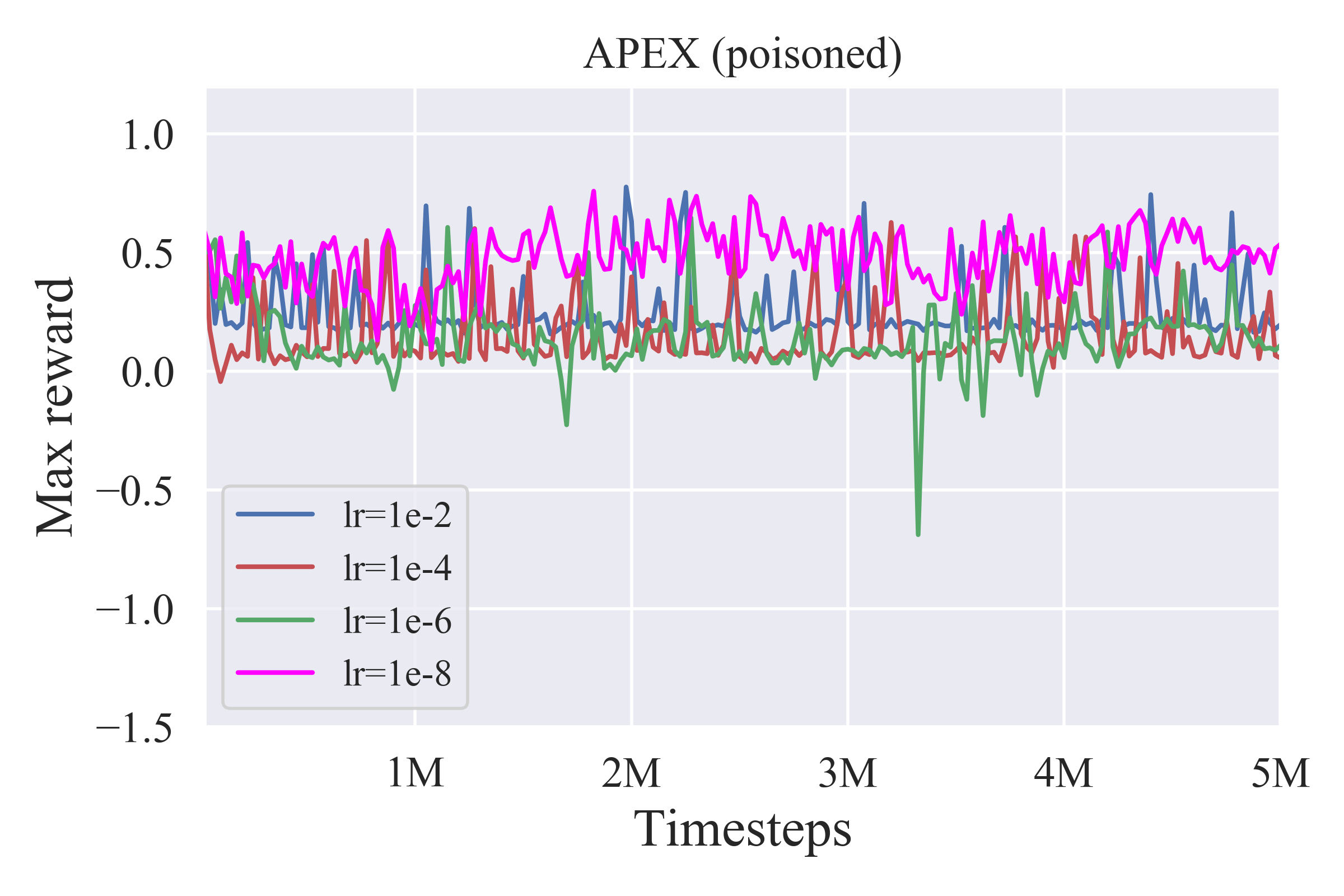} &
    \includegraphics[width=.3\textwidth]{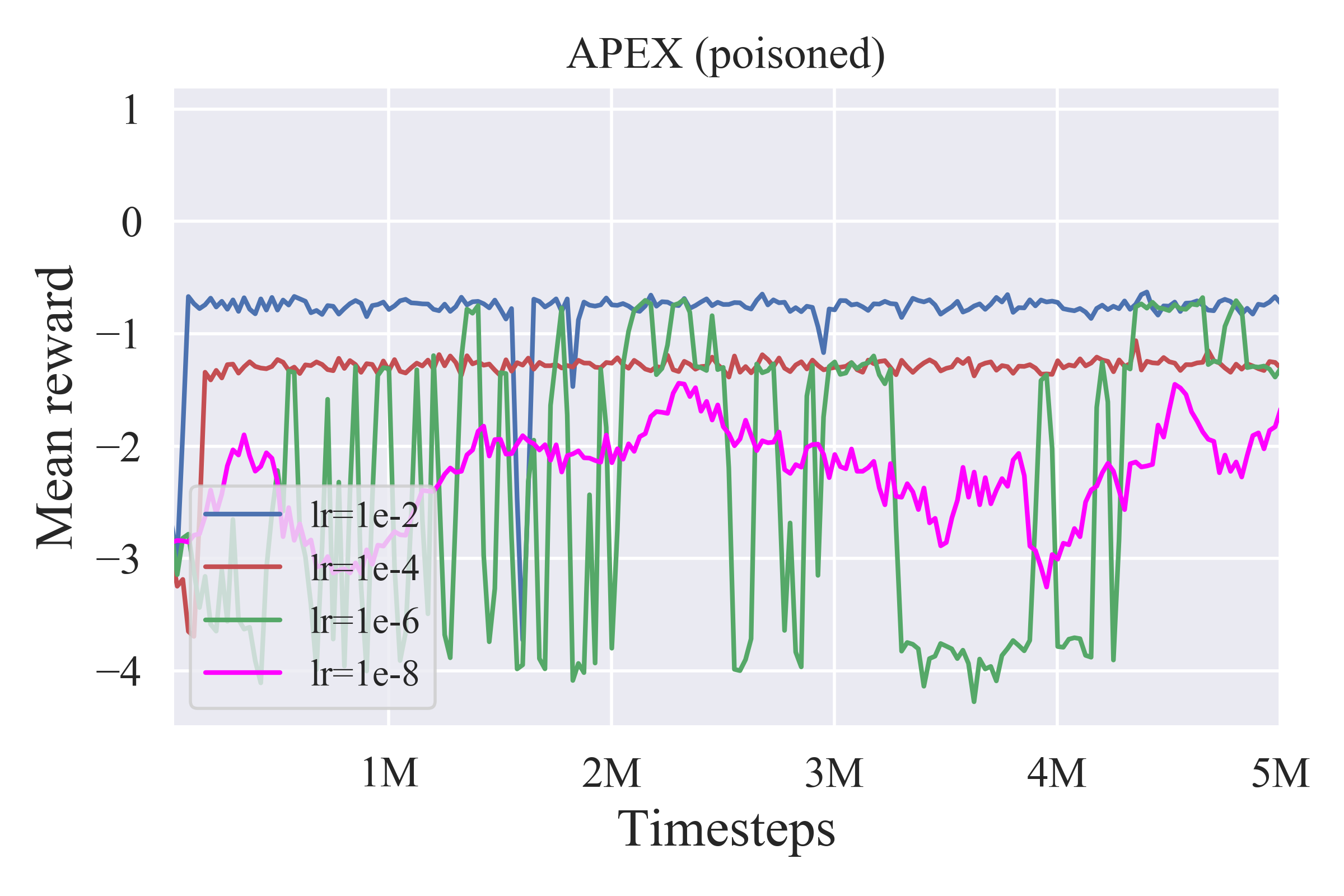} 
\end{tabular}
\caption{Blue agent's performance, learning with APEX-DQN with the same configuration as in Figure ~\ref{fig:traditional_attack}, except that the adversary performs an exfiltration attack with deception. The left plot shows that the red agent successfully exfiltrates the crown jewel most of the time. Furthermore the blue agent fails to learn a suitable policy.}
\label{fig:attacks}
\end{figure*}

Our previous results~\cite{molina-markham_network_2021} also demonstrated the
effect of allowing an adversary to deviate its behavior from the assumptions
outlined above with only gray-like actions. This simple deceptive behavior
degrades the performance of the blue agent (Figure~\ref{fig:attacks}). 

\section{Related Work}

Research has identified the need for gradually increasing the complexity of an
environment to make the learning process
feasible~\cite{openai_solving_2019,dennis_emergent_2020}. Moreover, Dennis et
al. propose an approach to find a curriculum of increasingly complex
environments~\cite{dennis_emergent_2020}. However, to our knowledge, we are the
first to propose a concrete approach to define distributions of network
environments for autonomous cyberdefense~\cite{molina-markham_network_2021}. Our
work also resulted in the development of FARLAND, to train network defenders
beyond what is possible with existing RL frameworks~\cite{brockman_openai_2016,
beattie_deepmind_2016, johnson_malmo_2016}.

Dozens of papers have explored RL for cybersecurity ~\cite{nguyen_deep_2019}.
However, prior work primarily addresses the problem: to what extent can RL
automate complex tasks? This has been attempted only in simple
scenarios without offering generalizable insights. Furthermore, prior efforts do
not properly address the question of how RL can be \emph{securely} applied to
solve complex tasks in the presence of adversaries. While many papers have
highlighted vulnerabilities of RL to adversarial manipulation
\cite{xiao_characterizing_2019, behzadan_faults_2018, kos_delving_2017,
lin_tactics_2017,han_reinforcement_2018,han_adversarial_2019,gleave_adversarial_2019},
there remains no fundamental understanding about how to secure RL agents during
learning and while making decisions. Many papers that have exposed RL’s
vulnerabilities to adversarial manipulation have done so under unrealistic
assumptions (e.g., assuming that observations and/or rewards can be directly
manipulated by adversaries~\cite{han_reinforcement_2018,kos_delving_2017}). We
argue that in the context of defending a network, such data manipulation
requires that an attacker tampers with traffic and client logs, which can be
prevented via traditional means, such as encryption. Our position is that
indirect observation manipulation (also known as environment
manipulation~\cite{xiao_characterizing_2019} or adversarial
policies~\cite{gleave_adversarial_2019}) and actuator
manipulation~\cite{behzadan_faults_2018} are more realistic threats. 

Securing RL agents is critical when they are deployed in adversarial situations
(e.g., using automation to defend an enterprise network). Existing threat
emulators~\cite{zilberman_sok_2020} do not support research to answer the
following types of questions: To what extent can we use RL to develop agents
that learn to perform security-related tasks? Furthermore, how do we measure the
robustness of such agents to deception (poisoning and evasion) attacks?
Previous~\cite{baillie_cyborg_2020} attempts to provide a framework to answer
the first type of questions. We argue that to address both questions, it is
necessary to model a diverse set of network environments and adversaries. 
\section{Conclusion}

Our position is that in order to develop and evaluate the next-generation
autonomous cyberdefenders it is necessary to model distributions of network
environments. We highlight how MITRE's approach (\emph{Network Environment
Design for Autonomous Cyberdefense}~\cite{molina-markham_network_2021})
leverages generative programs to model these network environment distributions.
Our work lists concrete features of network environments that researchers
must be able to parametrize to be able to make progress toward practical
autonomous cyberdefense. Importantly, our position accounts for the reality that
AI-enabled cyberdefenders will inevitably themselves be the targets of attacks
seeking to poison or evade their models. Our experimental results indicate this
looming threat. 

Our approach inspired the development of
FARLAND~\cite{molina-markham_network_2021}, used at MITRE to develop RL network
defenders that execute actions from the MITRE Shield
matrix~\cite{noauthor_mitre_2020} against attackers with TTPs that drift from
MITRE ATT\&CK TTPs~\cite{strom_mitre_2018}. While prior RL work has achieved
human-level or better performance in complex tasks (e.g., strategy games),
translating such achievements for cyberdefense requires a greater real-world
fidelity and flexibility to shifting rules and dramatically different
conditions.

We invite the research community to develop evaluation approaches based on the
performance of agents on distributions of network environments. Evaluations
based on the performance of agents on sets of predominantly fixed network
environments~--e.g., a list of red TTPs under similar network conditions~--are
less suited to reasoning about the performance of these agents when the
characteristics of adversarial TTPs, QoS requirements, and actions available to
the defender change. In particular, we caution against applying AI techniques,
such as RL, to tackle the network defense problem, without implementing robust
defenses against AI-targeted attacks.

\section*{Acknowledgements}

FARLAND is the result of contributions from researchers at MITRE and NSA,
including MITRE researchers Dr. Andres Molina-Markham, Dr. Ransom Winder, Cory
Miniter, Becky Powell, Bryan L. Quinn, Ceyer Wakilpoor, and Hunter DiCicco, and
NSA researcher Dr. Ahmad Ridley.

%-------------------------------------------------------------------------------
%\newpage

%\printbibliography[title=Bibliography]
\bibliographystyle{plain}
\bibliography{00_nd_not_a_game}

%%%%%%%%%%%%%%%%%%%%%%%%%%%%%%%%%%%%%%%%%%%%%%%%%%%%%%%%%%%%%%%%%%%%%%%%%%%%%%%%
\end{document}